\documentstyle[aps,prb,preprint]{revtex}
\begin{document}
\tightenlines
\draft
\title{Vortex Matter and its Phase Transitions}
\author{P. Chaddah and S. B. Roy}
\address{Low Temperature Physics Laboratory,\\
Centre for Advanced Technology,\\ Indore 452013, India}

\date{\today}
\maketitle
\begin{abstract}
The mixed state of type II superconductors has magnetic flux
penetrating the sample in the form of vortices, with each vortex
carrying an identical quantum of flux. These vortices generally
form a triangular lattice under weak mutually repulsive forces;
the lattice spacing can be easily varied over many orders of
magnitude by varying the external magnetic field. The elastic
moduli of this lattice are small and this soft vortex matter can
undergo phase transitions like normal matter, but with thermal
fluctuations and underlying defects playing an important role.
We discuss experimental studies on vortex matter phase
transitions, with some emphasis on DC magnetisation measurements
investigating the nature of the phase transition.
\end{abstract}



\section{Introduction}
All the technologically important superconductors, including the
so-called high-T$_{C}$ superconductors (HTSC), belong to the category
of type II superconductors.  The magnetic field (H) vs
temperature (T) phase diagram of these materials comprises two
distinct superconducting regions. As depicted in figure 1, the
upper critical field H$_{C2}$(T) provides the limiting boundary above
which superconductivity does not exist. Below the lower critical
field H$_{C1}$(T) we have the perfectly diamagnetic
Meissner-Oschenfeld phase. In the region within H$_{C1}$ and H$_{C2}$
magnetic flux penetrates the superconductor in the form of
vortices with each vortex carrying an identical quantum of flux.
Traditionally, these vortices were taken to have no structure
along their long axis parallel to the applied H. The
superconducting order parameter has a minimum at the centre of
each vortex, and the field has a maximum, as depicted in figure
2.  The field decays as one moves away from the vortex centre;
accordingly the vortices carry identical cylindrical current
shells at their periphery. These vortices repel one another as
expected by the identical current shells and Abrikosov had
argued that they form a triangular lattice. The lattice spacing
decreases continuously as the applied field is raised from H$_{C1}$
to H$_{C2}$. The dissipationless current-carrying applications of
superconductors require an understanding of vortex matter and
this has been continuously receiving attention \cite{1}.In this paper
we focus however on some studies regarding phase transitions
amongst different phases of vortex-matter. The density of vortex
matter can easily be varied over orders of magnitude, and
vortex-vortex interactions are weak. This makes vortex matter,
{\it a la colloids} \cite{2}, a potential testing ground for our understanding
of structural phase transitions.

\section{Brief history of early studies}

Questions regarding structural changes in vortex matter were
first addressed over three decades ago, soon after Abrikosov's
prediction of a triangular vortex lattice. We shall briefly
recapitulate those results as relevant today.  

Assuming the vortices to have no structure along their long
axis, the triangular vortex lattice will have only three
stiffness moduli.  These elastic constants become weak as H
approaches H$_{C2}$ because the modulation in local field, which
dictates the electromagnetic interaction between vortices, falls
towards zero. The interactions also tends to zero close to
H$_{C1}$ because the vortex separation rises. The elastic
constants of vortex matter are then many orders of magnitude
smaller than those of normal matter \cite{3}, and thermal
fluctuations as well as defects in the underlying
superconducting solid become important in the free energy of the
vortex solid. This can result in the vortex matter developing a
fluid-like structure close to H$_{C1}$ [Ref.4], as well as close
to H$_{C2}$ [Ref.5]. Labusch \cite{6} had shown that the shear
modulus goes parabolically to zero as the applied field nears
H$_{C2}$, and had concluded that {\it "This would indicate a
tendency of the fluxoid assembly to develop a fluid-like
character near H$_{C2}$ instead of forming a regular lattice"} .
Labusch also recognised \cite{6} the role of crystal
imperfections (in the underlying superconducting solid) in
whether the vortex matter would actually transform. Eilenberger
\cite{7} had earlier studied
thermodynamic fluctuations of the order parameter for H near H$_{C2}$
and had concluded {\it "the vortex lattice melts at a certain
external field strength below H$_{C2}$. This would be rather
interesting if it were observable experimentally."} His prophecy
proved right after nearly thirty years!  

For very low vortex
density near H$_{C1}$, the shear modulus again becomes small and a
fluid-like response was predicted \cite{4}. Trauble and Essmann, who had
pioneered the so-called decoration technique for imaging vortex
structures, have at least twice reported such interesting
observations. In their studies on Pb-(6 at \%)In, Trauble and
Essmann \cite{8} found that above a critical vortex lattice spacing of
d$_C$ = 8000 \AA (corresponding to a magnetic induction of about 4
mTesla) {\it "the flux-line pattern resembles an instantaneous
pattern of a two-dimensional fluid".} Similarly, in Pb-Tl Essmann
and Trauble \cite{9} observed a fluid-like structure for magnetic
induction below 2.5mTesla.

These early works on vortex matter
transforming from a solid to a fluid near both H$_{C1}$ and H$_{C2}$ are,
for reasons that are unclear, not much discussed in current
literature. This is not true of predictions made around the same
time by Fulde and Ferrel \cite{10}, and by Larkin and
Ovchinnikov \cite{11},
that in paramagnetic superconductors the superconducting order
parameter would develop modulations along the length of
vortices. This would occur at fields close to H$_{C2}$ and was
predicted \cite{12} to be a first order phase transition (FOPT) to the
Fulde-Ferrel-Larkin-Ovchinnikov (FFLO) state. The FFLO theory is
still used by theorists studying coexisting superconductivity
and weak magnetism \cite{13}. The prediction of a FOPT has motivated
experimentalists in recent years, as will be discussed later.

\section{Theoretical background}

\subsection{Possible phase transitions}

  We now briefly enumerate recent
theoretical ideas on possible phase transitions in stationary
vortex matter. It has been argued (see Blatter \cite{17} and references
therein) that a pure type II superconductor must show a
reentrant liquid phase as depicted in figure3.The triangular
vortex lattice is expected to melt with rising T via a FOPT. It
would also melt both as H is raised and as H is lowered.
Sophisticated theories have also argued that in the presence of
weak quenched disorder the low-H low-T phase would be a Bragg
glass (BG) where the position correlation function decays as a
power law. This BG phase of vortex matter would also melt with
increasing temperature via a FOPT. With increasing field there
would be a sudden proliferation of dislocation lines and the BG
phase would transform to a Vortex-glass (VG) phase. There has
been a recent theoretical prediction \cite{18} that this BG to VG
transition would also be a FOPT.  We shall discuss experiments
investigating the thermodynamic nature of these transitions in
the next section.

There has also been a lot of work on periodic
vortex lattices that are not triangular. Kogan and
collaborators \cite{19} have argued that non-local effects in the vortex
interaction can change the vortex lattice symmetry, to other
than triangular, as a function of H and T. Such changes have
been observed experimentally in the rare-earth nickel
borocarbides \cite{20}, in the HTSC material YBa$_2$Cu$_3$O$_7$
[Ref. 21], as well as in
V$_3$Si [Ref. 22] . This is presently another very active area, but we
shall not cover it in this paper because our emphasis is on
experimental studies concerning the thermodynamic nature of
vortex matter phase transitions.

Many developments of concepts
in phase transitions have also influenced recent studies on
vortex matter. Dynamic phase transitions have been proposed in
driven lattices \cite{14} with velocity being inversely related to an
effective temperature. Experimental studies have shown very
interesting results on driven vortex matter and its effective
phases (see Ref. [15,16] and references therein). This new and
interesting field is outside the scope of the present paper and
those interested can follow the references cited above.
  
\subsection{On signatures of first order phase transitions}


We first recall that a phase transition differs from the
colloquial 'transformation' in that a thermodynamic parameter
will change discontinuously when the phase boundary is crossed
by varying a control parameter like T or H. To report a FOPT
along a T$_{C}$(H) line, one needs to observe a discontinuous change
in entropy (i.e. observe a latent heat L), or in magnetisation
(i.e. in vortex volume), as one crosses the T$_{C}$(H) line by
varying either of the control variables T or H. The FOPT is
firmly established if the magnetisation jump M and L satisfy the
Clausius-Clapeyron relation; for this we also need dT$_{C}$/dH, and
thus need to locate T$_{C}$(H) over a finite region of H-T space.  If
we have only observed a discontinuous change in some
thermodynamic variable that is a second or higher derivative of
free energy (like susceptibility), then we can report a
(continuous) phase transition only.  If we have observed only
two regions of H-T space where the phases have widely differing
thermodynamic properties then more rigorous experimental checks
should be sought before we can even assert that the phase
transformation occurs through a phase transition.  

It is however
common (and justified) to report less rigorous signatures as
possible indicators of a FOPT. Amongst these are coexistence of
two phases, and supercooling/superheating across the T$_{C}$(H) line.
The latter feature presents itself as hysteresis when locating
the phase boundary (as a function of H or T) via a sharp change
in the measured property. To understand the need for
cross-checks after observing these indicators, we consider a
scenario \cite{23} in which hysteresis may be seen without a FOPT. Let
us say that we are pressure cycling a material which is very
viscous (or is an amorphous solid), and there is a pressure
range where the volume changes very strongly with pressure, but
not discontinuously. If we are changing pressure fast then the
observed volume is not the equilibrium volume, but lags behind
because of the high viscosity. A similar effect is expected when
pressure is being reduced. Thus we would observe a hysteresis
loop that is purely kinetic in origin. Kinetic hysteresis can be
seen when molecules in amorphous solids (or vortices in hard
superconductors) exhibit hindered kinetics, and equilibrium can
be reached only over times much longer than experimental time
scales. This is very relevant for vortex matter in hard
superconductors where M-H hysteresis is understood using Bean's
critical state model; which has hindered kinetics of vortices
(or pinning of vortices) as its cornerstone. Metastability due
to slow kinetics may also be possible at a second order phase
transition due to critical slowing down. Hysteresis and
metastability, by themselves, are thus arguable signatures of a
FOPT.

The T$_{C}$(H) line can be crossed by varying either of T or
H, and hysteresis in locating the FOPT can be seen with either
control variable. We have argued recently \cite{24} that for a FOPT the
range of supercooling (or extent of hysteresis) is more when T
is varied at constant H, than when H is varied at constant T.
This thus provides a necessary condition for hysteresis to be
attributed to a FOPT. A series of other predictions have been
made regarding metastabilities across a FOPT under different
experimental protocols \cite{24,25}. One that is relevant here is that
repeated cycling of magnetic field introduces a large
fluctuation energy that depends linearly on the number of
cycles, and nonlinearly on the amplitude of the cycling field.
These two parameters are also, usually, not strictly controlled
in AC measurements. Further kinetic hysteresis will be (more)
dominant in (higher frequency) AC measurements. We shall, in
this paper, discuss mainly DC measurements where both these
complications are minimal.

\section{Overview of experimental results}


\subsection{Vortex lattice melting}

As Eilenberger \cite{7} had prophesied, vortex lattice melting turned
out to interest many researchers. In YBa$_2$Cu$_3$O$_7$, all features of
a FOPT viz. hysteresis, vortex-volume discontinuity, and latent
heat have been observed and the Clausius-Clapeyron relation has
also been established. We summarize below a sequence of key
observations (in our opinion) that established this FOPT. 

The
resistivity of YBa$_2$Cu$_3$O$_7$ measured in a magnetic field showed a
sharp step while dropping to zero, and the resistivity step was
accompanied by hysteresis both with T and with H as the control
variable \cite{26,27,28}. This hysteresis was interpreted as a signature of
a FOPT, and ascribed to vortex lattice melting. Welp et al \cite{29}
stressed that transport measurements do not probe the defining
characteristics of a magnetic FOPT, and they showed that the
resistivity step was accompanied by a step in the magnetisation,
with the height of the step rising as the transition point moved
to lower T or higher H. The jump in magnetisation they observed,
however, occurred over a field range of the order of 0.1 Tesla.
This width has been understood as an effect of sample geometry
in these bulk measurements of magnetisation; Zeldov et al \cite{30}
showed in their studies on Bi$_2$Sr$_2$CaCu$_2$O$_8$ that the width becomes
negligible when local measurements are made using microhall
probes. Soon thereafter, Schilling et al \cite{31} measured the latent
heat across the transition in YBa$_2$Cu$_3$O$_7$, in conjunction with
measuring the jump in magnetisation.  Both these measurements
were made over a wide region of the melting line, and they
showed that over this entire region the Clausius-Clapeyron
relation was valid. Vortex lattice melting was thus firmly
established as a FOPT.  

True to Eilenberger's prophecy, vortex
lattice melting has since been investigated in many
superconductors and remains a very active area. Observation of
coexisting liquid and solid vortex matter has also been reported
recently \cite{32}. We should mention here that the phase diagram
depicted in figure 3, with reentrant melting, has also provoked
experimentalists . A reentrance of the peak-effect (in J$_C$ vs H),
that resembles the reentrance of the theoretically proposed
melting curve, has been observed in few samples of 2H-NbSe$_2$
[Ref. 33,34]. Questions regarding the influence of point defects, as
well as of extended line defects, on melting of vortex matter
also remain and the results will have some relevance to the
behaviour of normal matter (see e.g. Ref [35,36]).

\subsection{First order solid to solid phase transition}
 
The role of topological defects in producing an amorphous vortex
solid (as H is raised towards H$_{C2}$ ) has been discussed early in
conventional superconductors \cite{37}, and an amorphous solid has been
observed recently by neutron scattering\cite{38}. These studies have
not, however, addressed the nature of the phase transition (if
any) between the solid phases. While there are many reports of
such a transition in the HTSC compounds \cite{39,40,41}, both theory and
experiments have begun questioning the nature of the transition
(i.e. whether or not it is a FOPT) only very recently.
Experimentalists  motivated by the FFLO theory, which predicts a
solid-to-solid FOPT, were testing the onset of the PE for the
nature of the phase transition earlier, and we shall first
describe these experiments.  

In a generalised version of the
original FFLO theory, it was argued that the vortices are
segmented into short strings with concomitant enhancement of
pinning \cite{42,43}. This results in a peak-effect (PE) in J$_C$ vs H, or
in the M-H hysteresis curve, as H is raised towards H$_{C2}$ . CeRu$_2$
was a candidate material for an FFLO state ever since the
discovery of a magnetic anomaly near H$_{C2}$ [Ref. 44]. The first
thermodynamic signature indicating that the onset of PE is a
FOPT came through the observation that the PE starts at a field
H$^*_a$ on increasing field, but disappears at a lower field H$^*_d$ on
decreasing field \cite{45,46,47}. This hysteresis in the occurrence of the
PE was taken as the hysteresis expected in a FOPT. We have
attempted to identify other measurable signals of a FOPT, and
the FFLO theory motivated us, possibly as a red herring.We shall
now describe our DC magnetisation studies \cite{48,49,50} on CeRu$_2$, and
shall mention similar studies by other groups on CeRu$_2$ [Ref. 51,52],
NbSe$_2$ [Ref. 52-55] and YBa$_2$Cu$_3$O$_7$ [Ref. 55-57]. 
The motivating theories for the
studies on the last two materials were not FFLO but vortex
lattice softening and BG to VG transition.  

There has been no
reported observation so far of a latent heat associated with the
onset of PE in any of these materials. A step in the vortex
volume has been reported by us at the onset of PE in CeRu$_2$
[Ref. 48],
and by Ravikumar et al \cite{53} in NbSe$_2$. But these measurements are
tedious compared to similar measure for vortex lattice melting,
because the equilibrium magnetisation has to be extracted \cite{58} from
experimental M-H curves that are hysteretic. Since hysteresis in
locating the onset of PE had been reported as a signature of a
FOPT \cite{45,46,47}, it was natural to look for supercooled states; and
these were reported soon thereafter \cite{48,49}. Conventionally,
supercooling of a liquid is established by measuring diffusivity
which would decrease by orders of magnitude if the solid is
formed, but would fall smoothly as the liquid was supercooled
below the melting point. Since the FFLO state is characterised
by higher pinning or higher J$_C$, supercooling can be confirmed by
measuring minor hysteresis loops \cite{48,49,50} that are related to J$_C$
through the critical state model \cite{59}. [This technique of measuring
minor hysteresis loops (MHLs) has the added advantage that one
can ensure that hysteresis is due to bulk pinning \cite{60} (see figure
4). One can also ensure that no artefacts have been introduced
by the slight inhomogeneity of the magnetic field \cite{47}.] The two
phases on either side of the PE onset have different J$_C$ values,
and a supercooled metastable phase at a (H,T) point will produce
an MHL that is distinct from the MHL produced by the equilibrium
phase. The MHLs will thus be history-dependent in the region of
(H,T) space where supercooled metastable states can be made to
exist. This feature of history-dependent MHLs was observed by us
in various polycrystalline samples of CeRu$_2$ and was invoked by
us as evidence of supercooling and hence of a FOPT \cite{48,49,50}.
History-dependent MHLs have subsequently been observed in single
crystal samples of CeRu$_2$ [Ref. 51,52,61], and also in
NbSe$_2$ [Ref. 52-55] and YBa$_2$Cu$_3$O$_7$ 
[Ref. 55-57] single crystals. We must mention here that not
all samples of YBa$_2$Cu$_3$O$_7$ show history-dependant MHLs
\cite{57}.  

We were able to track this FOPT line in CeRu$_2$ over a factor of four in
vortex density, and the various histories under which MHLs were
observed corresponded to crossing this FOPT line through various
paths in (H,T) space. It was clear that supercooling persisted
farther when T is lowered at constant H, than when H is lowered
in constant T [Ref. 60]. This observation is consistent with the theory
for supercooling described in the previous section. This theory
makes many other predictions, one of which we now highlight. If
H$^*_{FC}$(T)is the limit to which supercooling can persist when T is
lowered in constant H, and H$^*_d$ (T) is the limit when H is
lowered at constant T, then it was predicted \cite{24,25} that since T$_{C}$
falls with rising H, the region separating H$^*_{FC}$(T) and H$^*_d$ (T)
will broaden as H rises (or as T falls). 
We have measured \cite{62} these
limits of supercooling in a single crystal sample of CeRu$_2$, and
the results are consistent with
predictions.  

As we had argued earlier, hysteresis can be
kinetic in origin and observation of hysteresis (or of phases
coexisting over finite time scales) may not be taken as a
sufficient indication of a FOPT. Na‹ve arguments \cite{25} suggest that
if hysteresis is kinetic in origin then the path dependence (in
H-T space) of the accompanying metastability would have an
inequality of sign opposite to that predicted for a FOPT. This
assumes some significance in the context of experiments to be
discussed below.  

We finally draw attention to two recent
experiments that conclude that the BG to VG transition in
Bi$_2$Sr$_2$CaCu$_2$O$_8$ is a FOPT. Gaifullin et al \cite{63}
 report a sharp change
in the plasma frequency at the BG-VG boundary and, since the
change here is at least as sharp as that across the melting
transition, they conclude that BG to VG is a FOPT. Using
magneto-optic studies, van der Beek et al \cite{64} observe coexistence
of the BG and VG phases, as well as supercooling of the VG
phase, and they also conclude that the BG to VG is a FOPT. We
must point out that they report coexistence for a few seconds
after a sudden lowering of the applied field, and this is not
seen to persist beyond about ten seconds. Going back to our
discussion on hindered kinetics (and kinetic hysteresis) in the
previous section, question arises whether van der Beek et al \cite{64} are
reporting a slow transient phenomenon. We then also recognise
that the measurements of Gaifullin et al \cite{63} are at high frequency.
In the view of the present authors, steady state measurements
are still necessary to establish the nature of the BG to VG
transition. We should mention here that we have earlier failed
to observe history-dependent MHLs in Bi$_2$Sr$_2$CaCu$_2$O$_8$ 
but ours were
DC measurements with a waiting time of about 100 seconds. It is
also possible that the supercooled state in Bi$_2$Sr$_2$CaCu$_2$O$_8$ is
very fragile and fluctuations induced during the measurement
process shattered the supercooled VG phase. The nature of the BG
to VG transition clearly needs more experiments, specially
looking for the extent of metastability after following
different paths in (H,T) space.

\section{Summary and conclusions}

We have discussed the current status of DC magnetisation studies
on phase transitions in stationary vortex matter. Vortex lattice
melting is established as a FOPT, while detailed experimental
studies show supercooling across a solid-to-solid transition of
vortex matter, indicating another FOPT. We have predicted a
specific path-dependence in the supercooling possible when T and
H (or vortex density) are varied across a FOPT, and the data on
CeRu$_2$ are consistent with this. While our predictions are valid
for normal matter like the water-to-ice transition, vortex
matter phase transitions are seen over much larger variations in
density and provide experimentally easier testing grounds.

\begin{figure}
\caption{The superconducting phase exists below the H$_{C2}$(T)
line. The perfectly diamagnetic Meissner-Oshenfeld phase exists
below the H$_{C1}$(T) line. The region between the H$_{C1}$(T) and
H$_{C2}$(T) lines is the mixed state where vortex matter
exists.}
\end{figure}
\begin{figure}
\caption{The order parameter       and the local induction B are plotted. The
vortices are a distance 'a' apart, and their centres (location is
indicated by the vertical lines) correspond to peaks in B and to the order
parameter dropping to zero. The amplitude of the oscillation in B falls as
H approaches H$_{C2}$.}
\end{figure}
\begin{figure}
\caption{Theoretically proposed phase diagram showing reentrant melting,
with vortex solid encompassed by the vortex liquid.}
\end{figure}
\begin{figure}
\caption{We show the behaviour of minor hysteresis loops drawn between the
envelope curves indicated by thicker lines. As indicated in the right
circle, the minor loops are straight lines if hysteresis is due to surface
currents. The left circle displays minor loops when hysteresis is due to
bulk pinning, as in critical state model. The minor loops are then
continuously nonlinear, breaking away from the straight line as soon as
they leave the envelope curves.}
\end{figure}
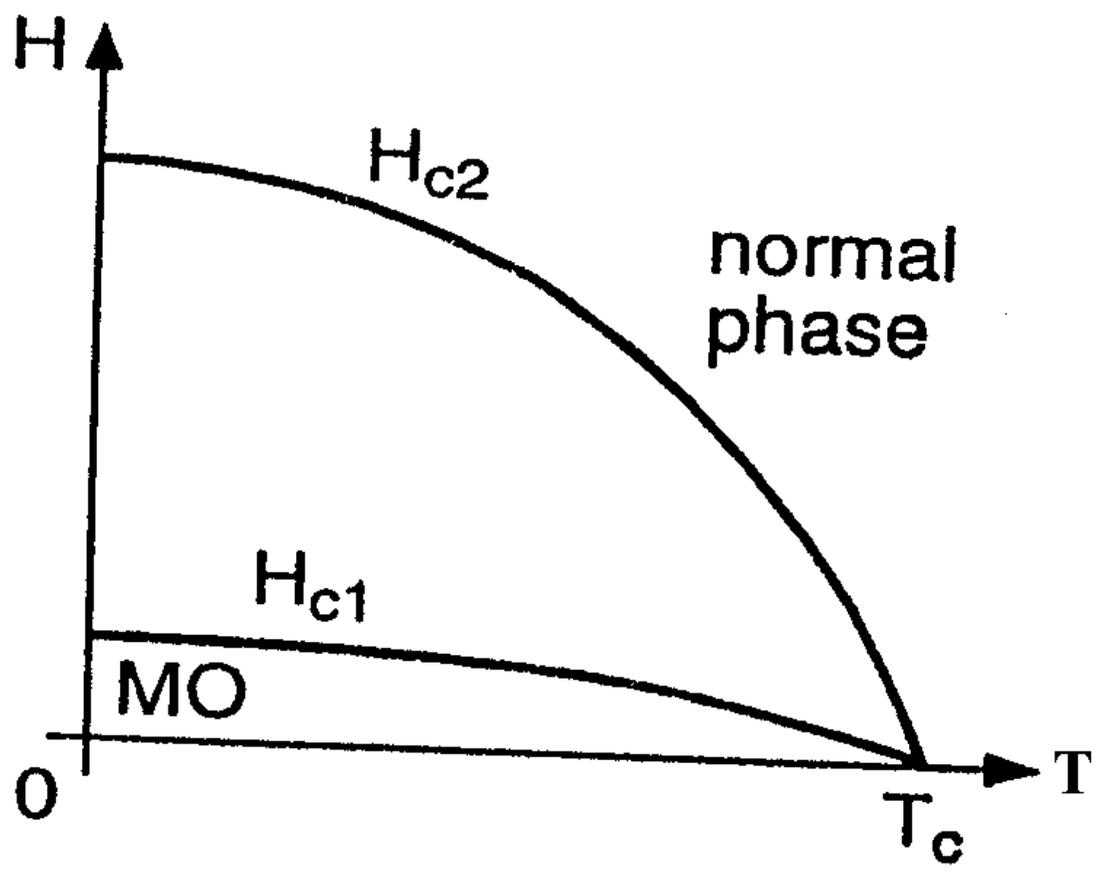
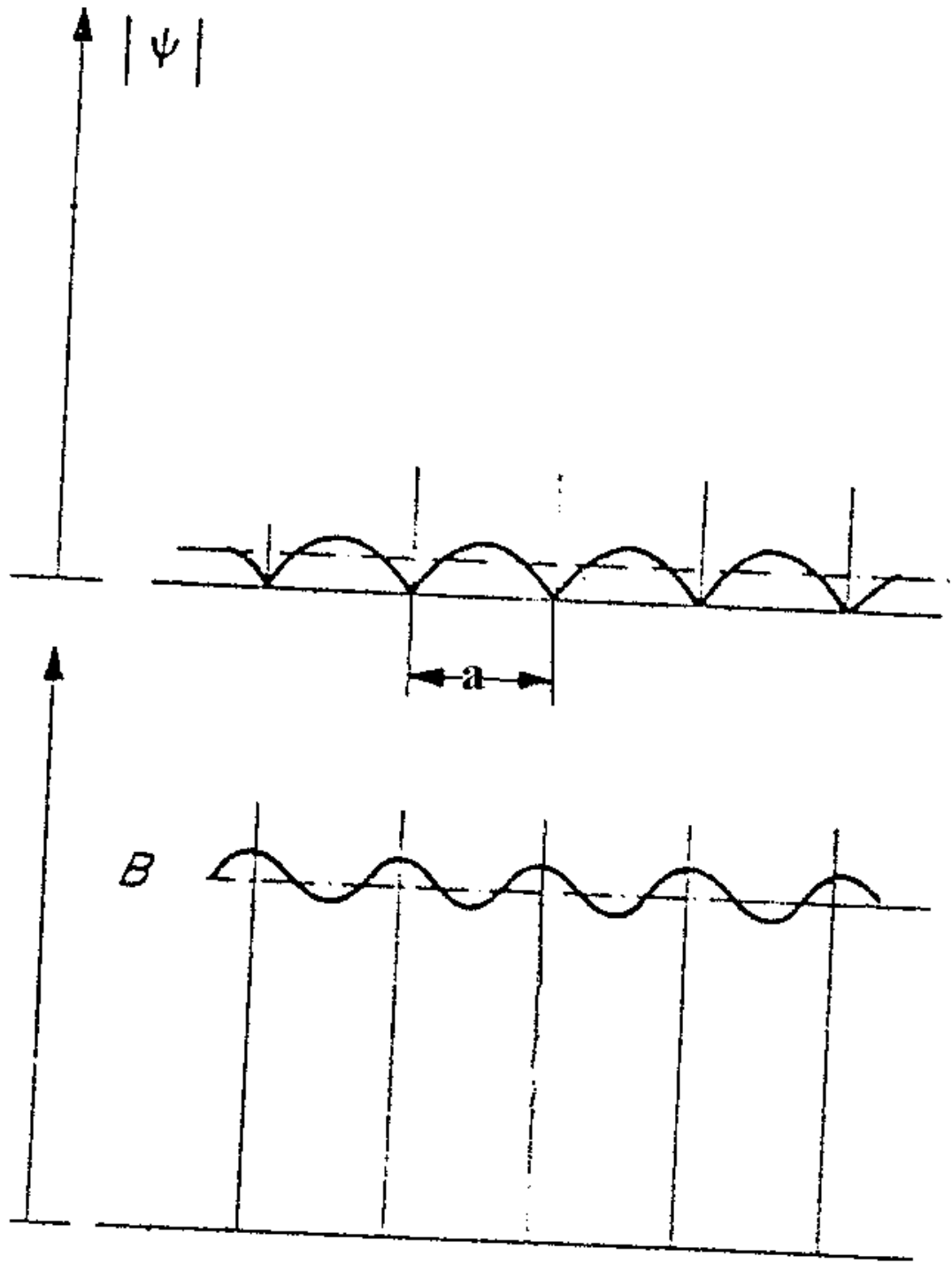
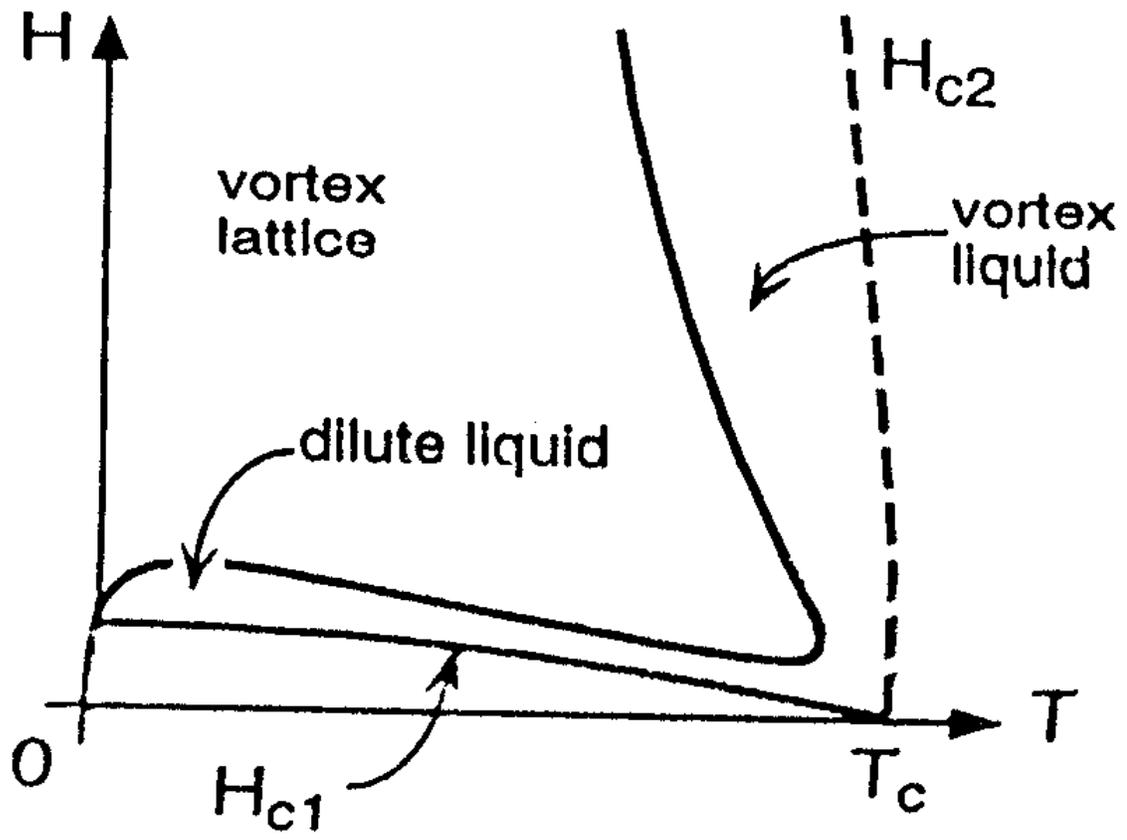
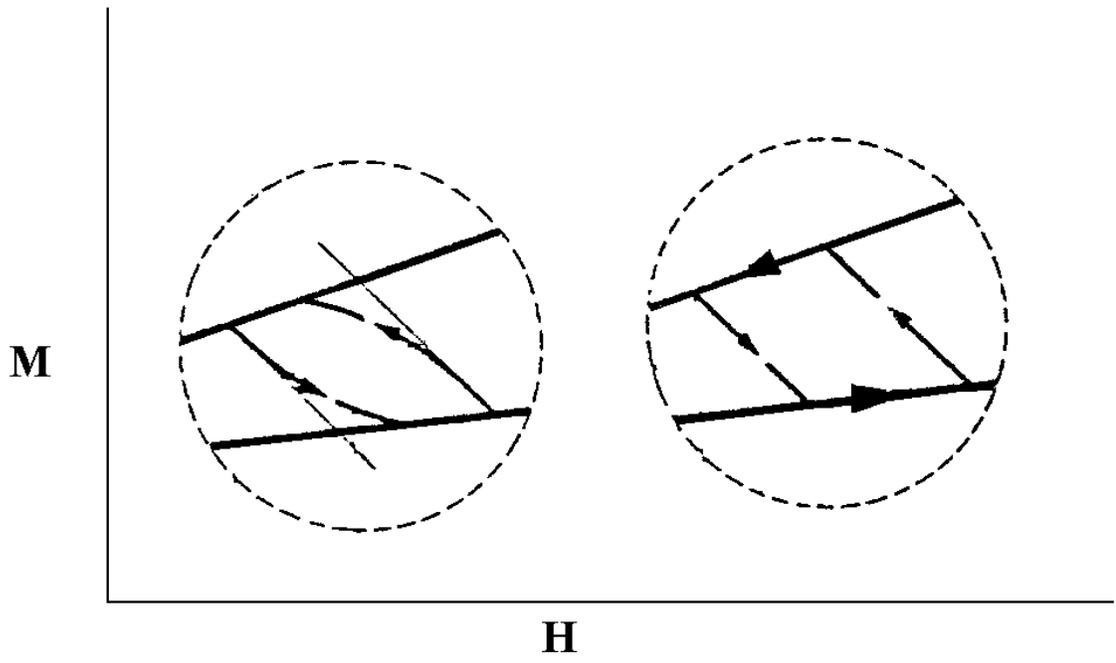
\end{document}